\documentclass[preprint]{ptephy_v1}
\begin{document}

\title{Spin-isospin excitation of ${}^3$He with three-proton final state}

\author{Souichi Ishikawa}
\affil{Science Research Center, 
                     Hosei University, 
                     2-17-1 Fujimi, Chiyoda, Tokyo 102-8160, 
                    Japan 
\email{ishikawa@hosei.ac.jp}
}





\begin{abstract}%
Spin-isospin excitation of ${}^3$He nucleus by proton-induced charge-exchange reaction, ${}^3\mathrm{He}(p,n)ppp$, at forward neutron scattering angle is studied in a plane wave impulse approximation (PWIA).  
In PWIA, cross sections of the reaction is written in terms of proton-neutron scattering amplitudes and  response functions of the transition from ${}^{3}$He to three-proton state by spin-isospin transition operators. 
The response functions are calculated with realistic nucleon-nucleon potential models using a Faddeev three-body  method. 
Calculated cross sections agree with available experimental data in substance.  
Possible effects arising from the uncertainty of proton-neutron amplitudes and three-nucleon interactions in three-proton system are examined.
\end{abstract}



\subjectindex{D00, D05, D22}

\maketitle



%
%
%
%


\section{Introduction}
\label{sec:intro}

Three-nucleon (3N) systems: ${}^3$H, ${}^3$He, nucleon-deuteron elastic and breakup reactions, etc., have been playing important roles in the quest for the details of interactions among nucleons.   
These systems are essentially total isospin $T=\frac{1}{2}$ states.
(Although the breaking of charge symmetry in nuclear interaction and the Coulomb interaction allow a mixture of $T=\frac{3}{2}$ components, its percentage is quite small \cite{Wu90}.)
On the other hand, knowledge of interactions among three nucleons, especially of three-nucleon interactions, in $T=\frac{3}{2}$ states is expected for studies on heavier nuclei, neutron-rich nuclei, neutron-star matter, etc.
Since there is no bound state of three-neutron ($3n$) and three-proton ($3p$) systems, which are typical $T=\frac{3}{2}$ states \cite{Pu10}, observables related these systems may be obtained from nuclear reactions that produce them as final continuum states.
Reaction mechanism of such reaction needs to be simple as possible to reduce ambiguity in extracting information on the nuclear interaction.

In the present paper, I will study a charge-exchange reaction: ${}^{3}\mathrm{He}(p,n)ppp$ reaction at incident proton energies of several hundreds MeV and the reaction angle $\theta_{n}=0^{\circ}$.
Although this is a four-body reaction that is still difficult to perform rigorous calculations at high energies, 
 the cross section of the reaction in PWIA is written in terms of $n(p,n)p$ ($pn$) scattering amplitudes and response functions of 3N system.
The former can be taken from nucleon-nucleon (NN) databases \cite{SAID,NN-online}.
The latter corresponds to a transition from the initial ${}^{3}$He bound state to final $3p$ continuum states, in which one needs to solve three-body problem.

The present author has developed a method to solve the quantum mechanical three-body problem applying the Faddeev method \cite{Fa61}.
This method is based on solving the Faddeev equation as integral equations in coordinate space, which even includes long range Coulomb force effects \cite{Sa79,Is03}, and has been successfully applied for the proton-deuteron systems \cite{Is09} and three alpha-particles systems \cite{Is13}.
In this paper, this method will be applied for calculating the response functions of $3p$ final states.

In Ref. \cite{Pa98}, the cross section $I(0^\circ)$ for the ${}^{3}\mathrm{He}(p,n)ppp$ reaction at the incident proton energy $T_p=200$ MeV was measured. 
In Ref. \cite{Wa08}, the polarization transfer coefficient in the transverse direction, $D_{NN}(0^\circ)$, and that in the longitudinal direction, $D_{LL}(0^\circ)$, as well as $I(0^\circ)$ were measured at $T_p=346$ MeV.
One of the measured polarization transfer coefficients, $D_{NN}(0^\circ)$, is consistent with the corresponding $pn$ values.
However, the other one, $D_{LL}(0^\circ)$, deviates from the $pn$ values.
The authors of Ref. \cite{Wa08} show that this discrepancy may be attributed  to a $3p$ resonance with spin-parity $\frac{1}{2}^{-}$.

Existence of resonant states in multi-neutron or multi-proton states has been a long-standing problem in nuclear physics. 
Recent compilation of the mass number $A=3$ systems \cite{Pu10} reports negatively for the existing of  $A=3$ resonance. 
In Ref. \cite{Ki16}, a possibility of existing of four-neutron (tetraneutron, $4n$) resonant state was reported. 
In Ref. \cite{Hi16}, it was shown that the existing of the $4n$ resonant state demands an attractive $T=\frac{3}{2}$ three-nucleon potential (3NP) that is tremendously strong.
Effects of such a 3NP on the $3p$ system will be studied.

In Sec. \ref{sec:formulation}, I will summarize the formalisms to calculate the response functions and then observables in  ${}^{3}\mathrm{He}(p,n)ppp$ $(\theta_{n}=0^\circ)$ reaction.
In Sec. \ref{sec:results}, I will show some results of calculations and compare them with available experimental data.
Summary will be given in Sec. \ref{seq:Summary}.
In appendix, some kinematical values related to the reaction will be summarized.

\section{Theoretical background}
\label{sec:formulation}

%
In this section, I will consider the charge-exchange reaction, ${}^{3}\mathrm{He}(\vec{p},\vec{n})ppp$ $(\theta_{n}=0^\circ)$ by PWIA, in which the $n(\vec{p},\vec{n})p$ $(\theta_{n}=0^\circ)$ scattering amplitude and response functions for the transition from ${}^{3}$He to $3p$ continuum states are the basic elements.  
 (See Refs. \cite{It94,Ic92}, e.g., for the general formalism of PWIA.)
Kinematics of the reaction is characterized by the incident proton energy in the laboratory (Lab.) system $T_p$, and the energy transfer in Lab. system $\omega_{Lab}$ defined by Eq. (\ref{eq:omega_lab}) in Appendix.
The direction of the incident proton and thus of the outgoing neutron is taken to be $z$-axis. 

First, I introduce the $3p$ Hamiltonian in the center of mass (c.m.) system,
\begin{equation}
\hat{H} _{3p} = \hat{H}_0 + \hat{V},
\end{equation}
where $\hat{H}_0$ is the kinetic energy operator of the three-body system, and $\hat{V}$ is an interaction potential, which consists of two-nucleon potentials (2NPs) and 3NPs.

Let $\vert \Psi_{m_1 m_2 m_3}^{(\pm)}(\boldsymbol{q},\boldsymbol{p})\rangle$ be an eigenstate of the Hamiltonian $\hat{H} _{3p}$ associated with an asymptotic $3p$-state, in which  the relative momentum between two protons is $\boldsymbol{q}$, the momentum of the third proton with respect to c.m. of the proton-pair  is $\boldsymbol{p}$, and the spin projection of the proton $i$ is $m_i$.
The superscript $(\pm)$ expresses the outgoing $(+)$ or incoming $(-)$  boundary condition.

The eigenvalue problems is written as 
\begin{equation}
\hat{H}_{3p} \vert  \Psi_{m_1 m_2 m_3}^{(\pm)} \left( \boldsymbol{q},\boldsymbol{p}\right) \rangle
= E(\boldsymbol{q},\boldsymbol{p})  \vert  \Psi_{m_1 m_2 m_3}^{(\pm)} \left( \boldsymbol{q},\boldsymbol{p}\right) \rangle,
\end{equation}
with 
\begin{equation}
E(\boldsymbol{q},\boldsymbol{p})  = \frac{q^2}{m_p}  + \frac{3 p^2}{4 m_p} ,
\end{equation}
where $m_p$ is the mass of the proton.

A response function corresponding to the transition from the ${}^3$He state with spin projection $M$, $\vert \Psi_M \rangle$, to $3p$-continuum states with energy $E$ by an operator $\hat{O}$ is given by
\begin{eqnarray}
R(E) &=& \frac{1}{2}\sum_{M=\pm\frac12} \sum_{m_1,m_2,m_3} \int d\boldsymbol{q} d\boldsymbol{p}
\left\vert T(\boldsymbol{q},\boldsymbol{p},m_1,m_2,m_3,M) \right\vert^2 
\delta\left(E-E(\boldsymbol{q},\boldsymbol{p})\right),
\label{eq:R_E}
\end{eqnarray}
where $E$ is related to kinematical values of the reaction as Eq. (\ref{Eq:E_3p}), and the transition amplitude is defined by 
\begin{equation}
T(\boldsymbol{q},\boldsymbol{p},m_1,m_2,m_3,M) = 
\langle  \Psi_{m_1 m_2 m_3}^{(-)} \left( \boldsymbol{q},\boldsymbol{p}\right) \vert \hat{O} \vert 
\Psi_M  \rangle.
\label{eq:T_amp}
\end{equation}

Using the completeness of the $3p$ states, we have
\begin{eqnarray}
R(E) &=& \frac{1}{2} \sum_{M=\pm\frac12}
\langle \Psi_{M} \vert\hat{O}^{\dagger} \delta(E-\hat{H}_{3p}) \hat{O}  \vert \Psi_{M} \rangle
\cr
&=&
-\frac{1}{2\pi} \sum_{M=\pm\frac12}  \mathrm{Im}  
\langle \Psi_{M} \vert\hat{O}^{\dagger}  \frac{1}{E+\imath \epsilon -\hat{H}_{3p}} \hat{O}  \vert \Psi_{M} \rangle.
\end{eqnarray}

Here, I introduce a wave function $\vert \Xi_M \rangle$ describing the disintegration process \cite{Is94},  
\begin{equation}
\vert \Xi_M \rangle = \frac{1}{E+\imath\epsilon - \hat{H}_{3p}} \hat{O} \vert \Psi_M \rangle, 
\label{eq:Psi_def}
\end{equation}
from which  the transition amplitude is calculated as follows:
\begin{equation}
T(\boldsymbol{q},\boldsymbol{p},m_1,m_2,m_3,M) = 
\langle  \Phi_{m_1,m_2,m_3}^{3p} \left( \boldsymbol{q},\boldsymbol{p}\right)\vert \hat{O} \vert  \Psi_M  \rangle
+ \langle  \Phi_{m_1,m_2,m_3}^{3p} \left( \boldsymbol{q},\boldsymbol{p}\right)\vert \hat{V} \vert  \Xi_M \rangle,
\label{eq:T_Xi}
\end{equation}
where $\vert \Phi_{m_1,m_2,m_3}^{3p} \left( \boldsymbol{q},\boldsymbol{p}\right) \rangle$ is the initial state corresponding to $\vert \Psi_{m_1,m_2,m_3}^{(+)} \left( \boldsymbol{q},\boldsymbol{p}\right) \rangle$.

Numerical solution of Eq. (\ref{eq:Psi_def}) is obtained by the method based on the Faddeev three-body theory \cite{Fa61}, whose formal and technical details are essentially same as those used for the proton-deuteron scattering \cite{Is03,Is09} and three alpha-particles \cite{Is13} problems.

The $n(\vec{p},\vec{n})p$ $(\theta_{n}=0^\circ)$ amplitude consists of  three independent terms:
\begin{equation}
f_{pn}
=
{\cal V}_c  +
 {\cal V}_L  \left(\boldsymbol{\sigma}_{p}\cdot\hat{\boldsymbol{z}}  \right) 
 \left(\boldsymbol{\sigma}_{n}\cdot\hat{\boldsymbol{z}}  \right) 
+
 {\cal V}_T\left(\boldsymbol{\sigma}_{p}\times\hat{\boldsymbol{z}}  \right) \cdot
 \left(\boldsymbol{\sigma}_{n}\times\hat{\boldsymbol{z}}  \right),  
\label{eq:f_i}
\end{equation}
where $\boldsymbol{\sigma}_{p}$ ($\boldsymbol{\sigma}_{n}$) is the Pauli spin matrix of the incident proton (the outgoing neutron); 
 ${\cal V}_c$, ${\cal V}_L$, and ${\cal V}_T$, are  spin-scalar,  spin-longitudinal, and  spin-transverse components of the amplitude, respectively.

The $pn$ observables, differential cross section, polarization transfer coefficients, are given as follows:
\begin{subequations}
\begin{equation}
\sigma^{pn}(0^\circ) =
\left\vert {\cal V}_c \right\vert^2 + \left\vert {\cal V}_L \right\vert^2 + 2 \left\vert {\cal V}_T \right\vert^2,
\label{eq:pn-obs-amp-a}
\end{equation}
\begin{equation}
D_{LL}^{pn}(0^\circ) = \frac{\left\vert {\cal V}_c \right\vert^2 + \left\vert {\cal V}_L  \right\vert^2 - 2 \left\vert {\cal V}_T \right\vert^2}{\left\vert {\cal V}_c  \right\vert^2 + \left\vert {\cal V}_L  \right\vert^2 + 2 \left\vert {\cal V}_T \right\vert^2},
\label{eq:pn-obs-amp-b}
\end{equation}
\begin{equation}
D_{NN}^{pn}(0^\circ) = \frac{\left\vert {\cal V}_c  \right\vert^2 - \left\vert {\cal V}_L  \right\vert^2 }{\left\vert {\cal V}_c  \right\vert^2 + \left\vert {\cal V}_L  \right\vert^2 + 2 \left\vert {\cal V}_T \right\vert^2}.
\label{eq:pn-obs-amp-c}
\end{equation}
\end{subequations}

In the process considered, there are three operators corresponding to each term of Eq. (\ref{eq:f_i}):
the isovector spin-scalar operator $\hat{O}_{c}$, 
the isovector spin-longitudinal operator $\hat{O}_{L}$, 
and the isovector spin-transverse operator $\hat{O}_{T}$, which are defined by
\begin{subequations}
\begin{equation}
\hat{O}_{c} = \sum_{i=1}^{3} e^{\imath Q_{c.m.} \hat{\boldsymbol{z}}\cdot \boldsymbol{r}_i}  t^{(+)}_{i},
\label{eq:O_c}
\end{equation}
\begin{equation}
\hat{O}_{L} = \sum_{i=1}^{3} e^{\imath Q_{c.m.} \hat{\boldsymbol{z}}\cdot \boldsymbol{r}_i}  \left(\hat{\boldsymbol{z}}\cdot\boldsymbol{\sigma}_{i}\right) t^{(+)}_{i},
\label{eq:O_L}
\end{equation}
\begin{equation}
\hat{O}_{T} = \sum_{i=1}^{3} e^{\imath Q_{c.m.} \hat{\boldsymbol{z}}\cdot \boldsymbol{r}_i}  \left(\hat{\boldsymbol{z}}\times\boldsymbol{\sigma}_{i}\right) t^{(+)}_{i},
\label{eq:O_T}
\end{equation}
\end{subequations}
where $Q_{c.m.}$ is the momentum transfer, Eq. (\ref{eq:Q_cm}) in Appendix, 
$t^{(+)}_{i}$  an isospin operator that transforms the neutron $i$ in ${}^3$He to proton $i$ in the final $3p$ state, 
$\boldsymbol{r}_i$ ($\boldsymbol{\sigma}_{i}$) the coordinate vector in the $3N$-c.m. system (the Pauli spin matrix) of the particle $i$.
The corresponding response functions will be denoted as $R_c(E)$,  $R_{L}(E)$, and  $R_{T}(E)$, respectively, 

The unpolarized differential cross section $I(0^\circ)$ and the polarization transfer coefficients, $D_{LL}(0^\circ)$ and $D_{NN}(0^\circ)$, for the ${}^{3}\mathrm{He}(p,n)ppp~(\theta_{n}=0^\circ)$ reaction are expressed as
\begin{subequations}
\begin{equation}
I(0^\circ) = N_{K}  
 \left(
\left\vert {\cal V}_c \right\vert^2 R_c+ \left\vert {\cal V}_L \right\vert^2 R_L + 2 \left\vert {\cal V}_T   \right\vert^2   R_T \right),
\label{eq:I_dcs}
\end{equation}
\begin{equation}
D_{LL}(0^\circ) = 
\frac
{\left\vert {\cal V}_c \right\vert^2 R_c + \left\vert {\cal V}_L \right\vert^2  R_L - 2 \left\vert {\cal V}_T\right\vert^2 R_T}
{ \left\vert {\cal V}_c \right\vert^2 R_c+ \left\vert {\cal V}_L \right\vert^2 R_L + 2 \left\vert {\cal V}_T   \right\vert^2   R_T},
\label{eq:D_LL}
\end{equation}
\begin{equation}
D_{NN}(0^\circ) = \frac{\left\vert {\cal V}_c \right\vert^2 R_c - \left\vert {\cal V}_L \right\vert^2 R_L }{ \left\vert {\cal V}_c \right\vert^2 R_c+ \left\vert {\cal V}_L \right\vert^2 R_L + 2 \left\vert {\cal V}_T   \right\vert^2   R_T},
\label{eq:D_NN}
\end{equation}
\end{subequations}
where a kinematical factor $N_{K}$ is given in Eq. (\ref{eq:N_K}) in Appendix.

\section{Results and discussion}
\label{sec:results}

In this section, calculations of the observables for the reactions, ${}^{3}\mathrm{He}(p,n)ppp$ $(\theta_{n}=0^\circ)$, at $T_p=346$ MeV and 200 MeV will be presented and compared with available   experimental data.

Calculations are performed as follows: 
the three-body equation, Eq. (\ref{eq:Psi_def}), is solved for each of the transition operators, Eqs. (\ref{eq:O_c}) - (\ref{eq:O_T}), from which the transition amplitude, Eq. (\ref{eq:T_amp}), is calculated by Eq. (\ref{eq:T_Xi}).
Then the response functions are calculated from Eq. (\ref{eq:R_E}).
Using the response functions together with the $pn$ amplitudes in Eq. (\ref{eq:f_i}), the observables are calculated by Eqs. (\ref{eq:pn-obs-amp-a}) - (\ref{eq:pn-obs-amp-c}).

In solving three-body equations, 3N partial wave states for which 2NPs and 3NPs are active, are restricted to those with total NN angular momenta $J \le 6$  for bound state calculations, and $J \le 4$ for continuum state calculations. 
For continuum state calculations, 3N states with total angular momenta $J_0 = \frac{1}{2}$  and $\frac{3}{2}$ are taken into account.
An error of these truncating procedures is estimated to be at most 2 \% from comparisons of results with $J \le 4$ NN states and those with $J \le 3$ ones,  and contributions from $J_0 = \frac{5}{2}$ states, which demonstrates that  it is good enough for the purposes of the present work.

As realistic models of 2NP,  the Argonne V$_{18}$ model (AV18) \cite{Wi95} and its V$_8$ version (AV8') \cite{Pu97}, the Argonne V$_{14}$ model (AV14) \cite{Wi84}, and a super-soft core model (dTRS)  \cite{dT75} are used. 
The NN scattering length parameters of these models for ${}^{1}\mathrm{S}_0$ states: $pp$, $nn$, and $pn$,  are compared  with empirical values \cite{Ma01} in Table \ref{tab:nn-pot}. 
As this table shows, AV8', AV14, and dTRS models are charge independent.
In this work, charge-dependent version of AV14 and dTRS in Table \ref{tab:nn-pot} are introduced by adding potentials that break the charge independence as done in Ref. \cite{Wu93}.
Such potentials for AV14 and dTRS are denoted by AV14(CD) and dTRS(CD), respectively.  

\begin{table}[t]
\caption{   
Empirical and calculated values for the NN scattering length parameters of ${}^{1}\mathrm{S}_0$- $pp$, $nn$, and $pn$ states.
For $pp$ system, the scattering length after subtracting the effect of Coulomb force is used.
Experimental values are taken from Ref. \cite{Ma01}.
}
\label{tab:nn-pot} 
\centering
\begin{tabular}{ccccc}
\hline
 & $a^{N}_{pp}$ (fm) & $a_{nn}$ (fm) & $a_{pn}$ (fm)  \\  
\hline
Empirical  & $-17.3\pm0.4$ & $-18.9\pm0.4$ &  $-23.740\pm0.020$\\
AV18& -16.6 & -18.3 &  -23.7  \\  
AV8'  & -19.3 & -19.3 & -19.3  \\ 
AV14 & -23.7 & -23.7 & -23.7 \\  
AV14 (CD) & -17.7 & -18.9 & -23.7 \\  
dTRS & -18.0 & -18.0 & -18.0 \\ 
dTRS (CD) & -16.6 & -18.0 & -24.1 \\  
\hline
\end{tabular}
\end{table}

The ${}^{3}$He wave function is calculated using each 2NP model with the Brazil model of the two-pion exchange three-nucleon potential given in Ref. \cite{Is07a}, whose cutoff mass parameter of the $\pi NN$-vertex $\Lambda_{\pi}$ is tuned to reproduce the empirical binding energy \cite{Is09}.
The values of $\Lambda_{\pi}$  in the unit of MeV are 660, 610, 670, and 650 for AV18, AV8', AV14(CD), and dTRS(CD), respectively.
It is noted that the ${}^{3}$He wave function by the CD version of  AV14 (dTRS) is used for calculations of the original version of  AV14 (dTRS).

The $n(\vec{p},\vec{n})p~(\theta_n=0^{\circ})$ scattering amplitudes in Eq. (\ref{eq:f_i}), ${\cal V}_c$, ${\cal V}_L$, and ${\cal V}_T$, are calculated by Eqs. (\ref{eq:pn-obs-amp-a}) - (\ref{eq:pn-obs-amp-c}) with the $pn$ observables, $\sigma^{pn}(0^\circ)$, $D_{LL}^{pn}(0^\circ)$, and $D_{NN}^{pn}(0^\circ)$, 
 taken from the SP07 solution \cite{SAID,Ar07}, which are shown in Table \ref{tab:coefs}.

\begin{table}[t]
\caption{   
Observables and scattering amplitudes in $n(\vec{p},\vec{n})p$ reaction at forward angle $\theta_n=0^\circ$ taken from the SP07 solution \cite{SAID,Ar07}.
Those used for the calculations of ${}^{3}\mathrm{He}(p,n)ppp$ reaction at $T_p=200$ MeV and 346 MeV are shown.
}
\label{tab:coefs} 
\centering
\begin{tabular}{cccc}
\hline
 & $T_p = 200$ MeV & $T_p=346$ MeV &  \\ 
\hline
$\sigma^{pn}(0^\circ)$  [mb/sr] & 12.47 & 11.32 \\
$D_{LL}^{pn}(0^\circ)$ & -0.1831 & -0.3942 \\
$D_{NN}^{pn}(0^\circ)$ & -0.3269& -0.2396 \\
$\vert{\cal V}_c \vert^2$ [mb/sr] & 0.5085  &0.3583 \\
$\vert{\cal V}_L \vert^2$ [mb/sr] & 4.5849 & 3.0705 \\
$\vert{\cal V}_T \vert^2$ [mb/sr] & 3.6883 & 3.9456 \\
\hline
\end{tabular}
\end{table}

\begin{figure}
\begin{center}
\includegraphics[width=0.4\columnwidth]{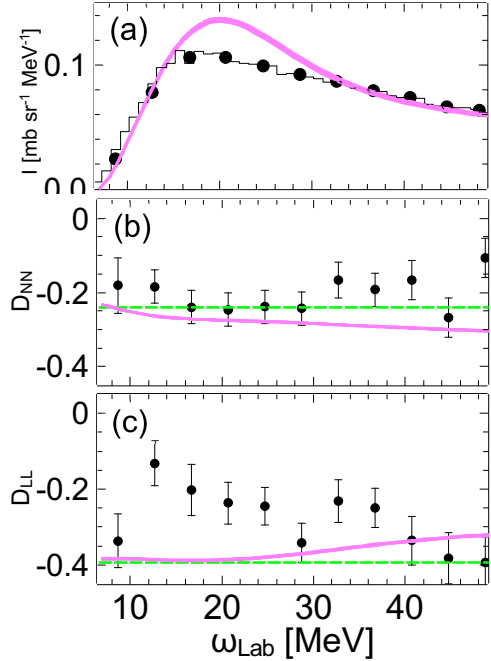}
\caption{
Differential cross section $I(0^\circ)$ (a) and  polarization transfer coefficients, $D_{NN}(0^\circ)$ (b) and $D_{LL}(0^\circ)$ (c), for the ${}^{3}\mathrm{He}(p,n)ppp$ reaction at $T_p=346$ MeV.
Calculations with all 2NP models in Table \ref{tab:nn-pot} are shown by bands (light magenta).
The experimental data (black points and histogram) are taken from Ref. \cite{Wa08}.
Dashed horizontal lines (green) in (b) and (c) are the corresponding $pn$ values in Table \ref{tab:coefs}. 
\label{fig:3He-pn-sig-DLL-DNN-2np}
}
\end{center}
\end{figure}

\begin{figure}
\begin{center}
\includegraphics[width=0.4\columnwidth,angle=-90]{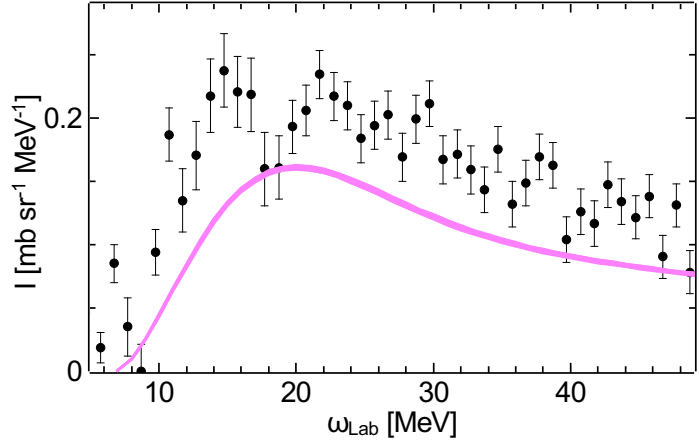}
\caption{
Differential cross section $I(0^\circ)$ for the ${}^{3}\mathrm{He}(p,n)ppp$ reaction at $T_p=200$ MeV.
Calculations with all 2NP models in Table \ref{tab:nn-pot} are shown by band (light magenta).
The experimental data are taken from Ref. \cite{Pa98}.
\label{fig:sigma-200MeV-2np}
}
\end{center}
\end{figure}

In Fig. \ref{fig:3He-pn-sig-DLL-DNN-2np}, calculated differential cross section $I(0^\circ)$ and  polarization transfer coefficients, $D_{NN}(0^\circ)$ and $D_{LL}(0^\circ)$, for the ${}^{3}\mathrm{He}(p,n)ppp$ reaction at  $T_p = 346$ MeV as functions of $\omega_{Lab},$ are compared with the experimental data of Ref. \cite{Wa08}.
In Fig. \ref{fig:sigma-200MeV-2np}, calculated values of $I(0^\circ)$ for $T_p=200$ MeV are compared with the experimental data \cite{Pa98}.
In both figures, calculations of all 2NP models in Table \ref{tab:coefs} fall within narrow bands, which demonstrates small $pp$-2NP dependency of the observables.

The calculations of $I(0^\circ)$ at $T_p=346$ MeV reproduce the data well except some deviations around $\omega_{Lab}=20$ MeV.
Those at $T_p=200$ MeV reproduce the lineshape of the data with a reduction by about 30\%. 

The calculated and experimental values of $D_{NN}(0^\circ)$ and calculated $D_{LL}(0^\circ)$ are almost consistent with the $pn$ values, which are expressed by the dashed horizontal lines in Figs. \ref{fig:3He-pn-sig-DLL-DNN-2np} (b) and (c).
On the other hand, the experimental values of $D_{LL}(0^\circ)$ deviate from the $pn$ values with an energy-transfer dependence, from which the authors of Ref. \cite{Wa08} predicted the existence of a $3p$ resonance in $\frac{1}{2}^{-}$ state centered at $\omega_{r}=16 \pm 1$ MeV with the width of $\Gamma = 11 \pm 3$ MeV.

\begin{figure}
\begin{center}
\includegraphics[width=0.4\columnwidth,angle=-90]{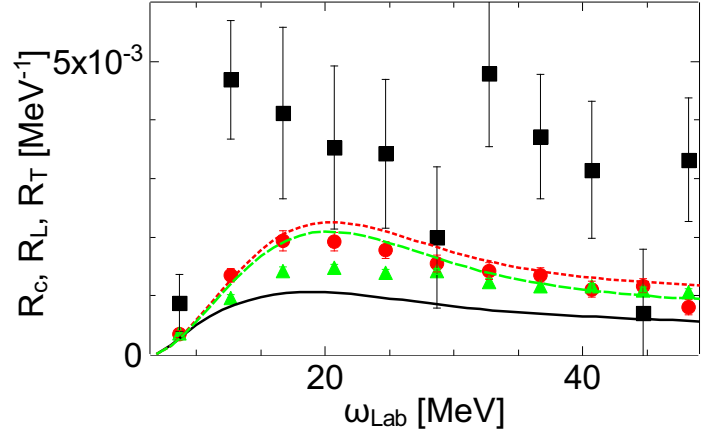}
\caption{
Data points with error bars are the response functions $R_c$ (black squares), $R_L$ (red circles), and $R_T$ (green triangles), extracted from the data \cite{Wa08} and the $pn$ amplitudes from the SP07 solution.
Curves are calculated  response functions with AV18 for $R_c$ (black solid line), $R_L$ (red dotted line), and $R_T$ (green dashed line).
\label{fig:Rc-RL-RT-exp}
}
\end{center}
\end{figure}

Using three observables measured in Ref. \cite{Wa08}  along with the $pn$ amplitudes in Table \ref{tab:coefs}, the response functions, $R_c$, $R_L$, and $R_T$, are calculated by Eqs. (\ref{eq:I_dcs}) - (\ref{eq:D_NN}). 
Thus obtained response functions are compared with the calculated ones with AV18 in Fig. \ref{fig:Rc-RL-RT-exp}. 
The figure shows that the extracted $R_L$ and $R_T$ have similar shape and magnitude as the calculations, but the extracted $R_c$ is a few times larger than the calculation. 
The resonance-like behavior in $D_{LL}(0^\circ)$ as a function of $\omega_{Lab}$ is reflected in $R_c$, but not in $R_L$ and $R_T$.
However, the calculations are not able to reproduce this tendency.

The observables in this work largely owe to  the $pn$ amplitudes, which are related $pn$ observables as Eqs. (\ref{eq:pn-obs-amp-a})-(\ref{eq:pn-obs-amp-c}).
Since there is few experimental data of corresponding $pn$ observables \cite{Ar00a},  the uncertainty in the $pn$ amplitudes used in this work is not small.
Thus, I have evaluated the $pn$ amplitudes inversely from  the calculated response functions,  $R_c$, $R_L$, and $R_T$, and the experimental data of $I(0^\circ)$, $D_{NN}(0^\circ)$, and $D_{LL}(0^\circ)$ by Eqs. (\ref{eq:I_dcs}) - (\ref{eq:D_NN}).

The $pn$-amplitudes obtained in this way depend on $\omega_{Lab}$ that the experimental data exist.
Taking the average, one obtains: 
$\vert {\cal V}_c \vert^2 = 1.43 \pm 0.57$~mb/sr, 
$\vert {\cal V}_L \vert^2 = 2.90 \pm 0.38$~mb/sr,
and
$\vert {\cal V}_T \vert^2 = 3.60 \pm 0.75$~mb/sr, from which the $pn$ observables are calculated as:
$\sigma^{pn}(0^\circ)=11.5 \pm 1.7$~mb/sr,
$D_{LL}^{pn}(0^\circ)=-0.24 \pm 0.11$, and
$D_{TT}^{pn}(0^\circ)=-0.13 \pm 0.05$.
The errors are only statistical ones with respect to the averaging procedure, and effects of the experimental error are not taken into account.

\begin{figure}
\begin{center}
\includegraphics[width=0.4\columnwidth]{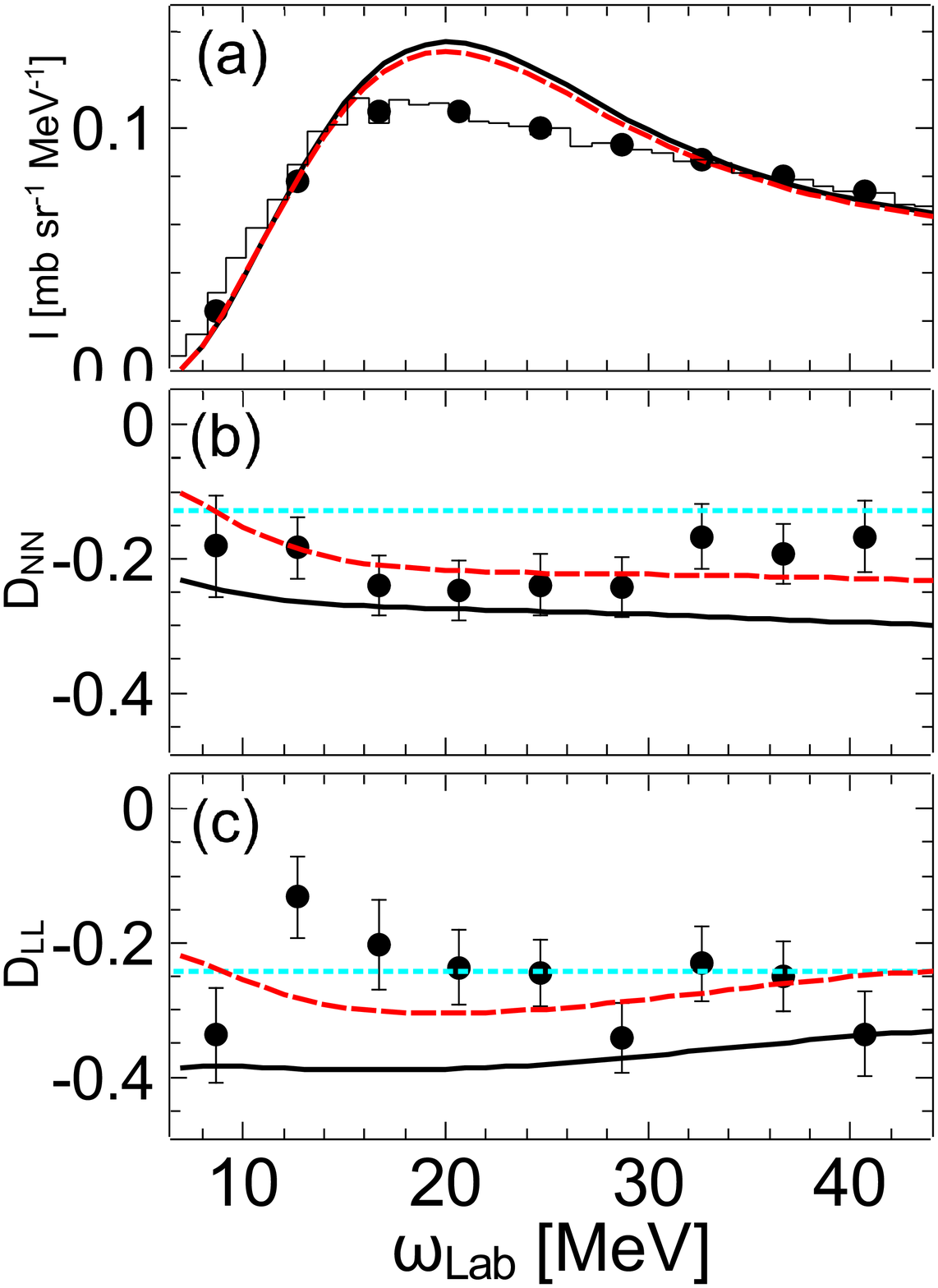}
\caption{
Same as Fig. \ref{fig:3He-pn-sig-DLL-DNN-2np}. Calculations for AV18 with the $pn$ amplitudes taken from the SP07 solution are shown by sold (black) lines.  
Those with the fitted $pn$ amplitudes (see the text) are shown by the dashed (red) lines.
Dotted horizontal lines (cyan) in (b) and (c) are the $pn$ values calculated with the fitted $pn$ amplitudes. 
\label{fig:3He-pn-sig-DLL-DNN-coefs} 
}
\end{center}
\end{figure}

Fig. \ref{fig:3He-pn-sig-DLL-DNN-coefs} shows the calculated observables with the above fitted $pn$-amplitudes (only the central values are used.), which gives a better agreement with the data, although the rapid dependence of $D_{LL}(0^\circ)$ is not reproduced.

Next, I will study effects of 3NP on the observables. 
Recently, the possibility of a resonant $4n$ state at low energy is indicated experimentally in Ref. \cite{Ki16}.
In Ref. \cite{Hi16}, effects of $T=\frac{3}{2}$ 3NP on $4n$-system as well as $3n$-system are studied. 
The functional form of 3NP used in Ref. \cite{Hi16} is as follows. 
\begin{equation}
V^{\mathrm{3NP}}(T) = \sum_{n=1}^{2} W_n(T) e^{-(r_{ij}^2+r_{jk}^2+r_{ki}^2)/b_n^2} {\cal P}_{ijk}(T), 
\end{equation}
where $T=\frac{1}{2}$ or $\frac{3}{2}$, $r_{ij}$ is the distance between the $i$-th and $j$-th nucleons, and ${\cal P}_{ijk}(T)$ is a projection operator on the 3N isospin $T$ state.

The range parameters used in Refs. \cite{Hi04,Hi16} are $b_1=4.0$ fm and $b_2=0.75$ fm.
The strength parameters of the shorter range term $W_2(T)$ for both of $T=\frac{1}{2}$ and $T=\frac{3}{2}$ are fixed to be $35.0$ MeV in Ref. \cite{Hi16}, and also in this work. 
The required value of the strength parameter for the longer range term $W_1(\frac{3}{2})$ for $J^\pi=0^+$ $4n$-state to bind as the lower bound of the experimental value \cite{Ki16} is $-36.14$ MeV \cite{Hi16}.
This value contrasts with $W_1(\frac{1}{2})=-2.04$ MeV, which is determined to reproduce the binding energies of ${}^{3}$H, ${}^{3}$He, and  ${}^{4}$He in combination with Argonne V$_8$' (AV8') NN potential \cite{Pu97}. 

In the following,  I will use the AV18, which is more repulsive than AV8' in 3N$(T=\frac{1}{2})$ bound state.
As a consequence of this, a more attractive value:  $W_1(\frac{1}{2})=-2.55$ MeV, is used to reproduce ${}^{3}$He binding energy.
However, this difference may not be essential in the present case.

\begin{figure}
\begin{center}
\includegraphics[width=0.4\columnwidth]{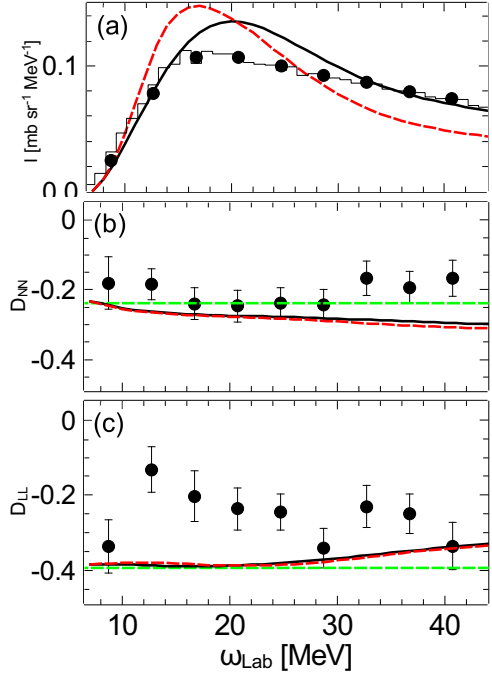}
\caption{
Same as Fig. \ref{fig:3He-pn-sig-DLL-DNN-2np}.
Calculations with AV18 are shown by sold (black) lines.  
Calculations with AV18 plus $V^{\mathrm{3NP}}(\frac{3}{2})$ taking $W_1(\frac{3}{2})=-36.0$ MeV are shown by dashed (red) lines. 
\label{fig:3He-pn-sig-DLL-DNN-m036}
}
\end{center}
\end{figure}

In Fig. \ref{fig:3He-pn-sig-DLL-DNN-m036}, calculated values with $V^{\mathrm{3NP}}(\frac{3}{2})$ taking $W_1(\frac{3}{2})=-36.0$ MeV are compared with the AV18 calculations.
The introduction of the $V^{\mathrm{3NP}}(\frac{3}{2})$ shifts the peak of the cross section to higher in the magnitude and lower in the position, which makes the agreement with the experimental data worse than the AV18 calculation. 
On the other hand, effects of the 3NP on the $D_{NN}(0^\circ)$ and $D_{LL}(0^\circ)$ are quite small.
These rather small effects of the 3NP on the $3p$ system in spite of the large value of $W_1(\frac{3}{2})$ are   however  consistent with the analysis of  $3n$ systems in Ref. \cite{Hi16}, and should be due to a large separation among three protons by the Pauli principle.

In Ref. \cite{Hi16}, dependence of the strength parameters in the 3NP on the total angular momentum and parity $J_0^\pi$ is not considered for simplicity.  
Here, I will examine the $J_0^\pi$-dependence of  the parameter $W_1(\frac{3}{2})$.
In Table \ref{tab:DLL-jp}, results of $D_{LL}(0^\circ)$ calculated including $V^{\mathrm{3NP}}(\frac{3}{2})$ with $W_1(\frac{3}{2})=-36.0$ MeV for all four states: $J_0^\pi = {\frac{1}{2}}^{\pm}$ and ${\frac{3}{2}}^{\pm}$,  or for only one partial wave state, are shown.
Even though the effects are not so large compared to the difference between the data and AV18 calculation, it looks that only $V^{\mathrm{3NP}}(\frac{3}{2})$ with $J_0^\pi = \frac{1}{2}^-$ is effective to the difference.

\begin{table}[t]
\caption{   
Calculated values of the polarization transfer coefficient $D_{LL}(0^{\circ})$ for the ${}^{3}\mathrm{He}(p,n)ppp$ reaction at $T_p=346$ MeV and at $\omega_{Lab}=15$ MeV including $V^{\mathrm{3NP}}(\frac{3}{2})$ with $W_1(\frac{3}{2})=-36.0$ MeV for all four states: $J_0^\pi = {\frac{1}{2}}^{\pm}$ and ${\frac{3}{2}}^{\pm}$,  or for only one partial wave state.
$\Delta D_{LL}^{pn}(0^\circ)$ is the difference from the AV18 calculation.
}
\label{tab:DLL-jp} 
\centering
\begin{tabular}{ccc}
\hline
 & $D_{LL}^{pn}(0^\circ)$ & $\Delta D_{LL}^{pn}(0^\circ)$ \\  
\hline
AV18   & -0.389 & \\
AV18+$V^{\mathrm{3NP}}(T=\frac{3}{2})$ $[J_0^\pi=\frac{1}{2}^{\pm},\frac{3}{2}^{\pm}]$ & -0.382 &0.007 \\
AV18+$V^{\mathrm{3NP}}(T=\frac{3}{2})$ $[J_0^\pi=\frac{1}{2}^{+}]$ & -0.391  & -0.002 \\
AV18+$V^{\mathrm{3NP}}(T=\frac{3}{2})$ $[J_0^\pi=\frac{1}{2}^{-}]$ & -0.379 &0.010 \\
AV18+$V^{\mathrm{3NP}}(T=\frac{3}{2})$ $[J_0^\pi=\frac{3}{2}^{+}]$ & -0.389  & 0.000\\
AV18+$V^{\mathrm{3NP}}(T=\frac{3}{2})$ $[J_0^\pi=\frac{3}{2}^{-}]$ & -0.388  & 0.001\\
\hline
\end{tabular}
\end{table}

\begin{figure}
\begin{center}
\includegraphics[width=0.4\columnwidth]{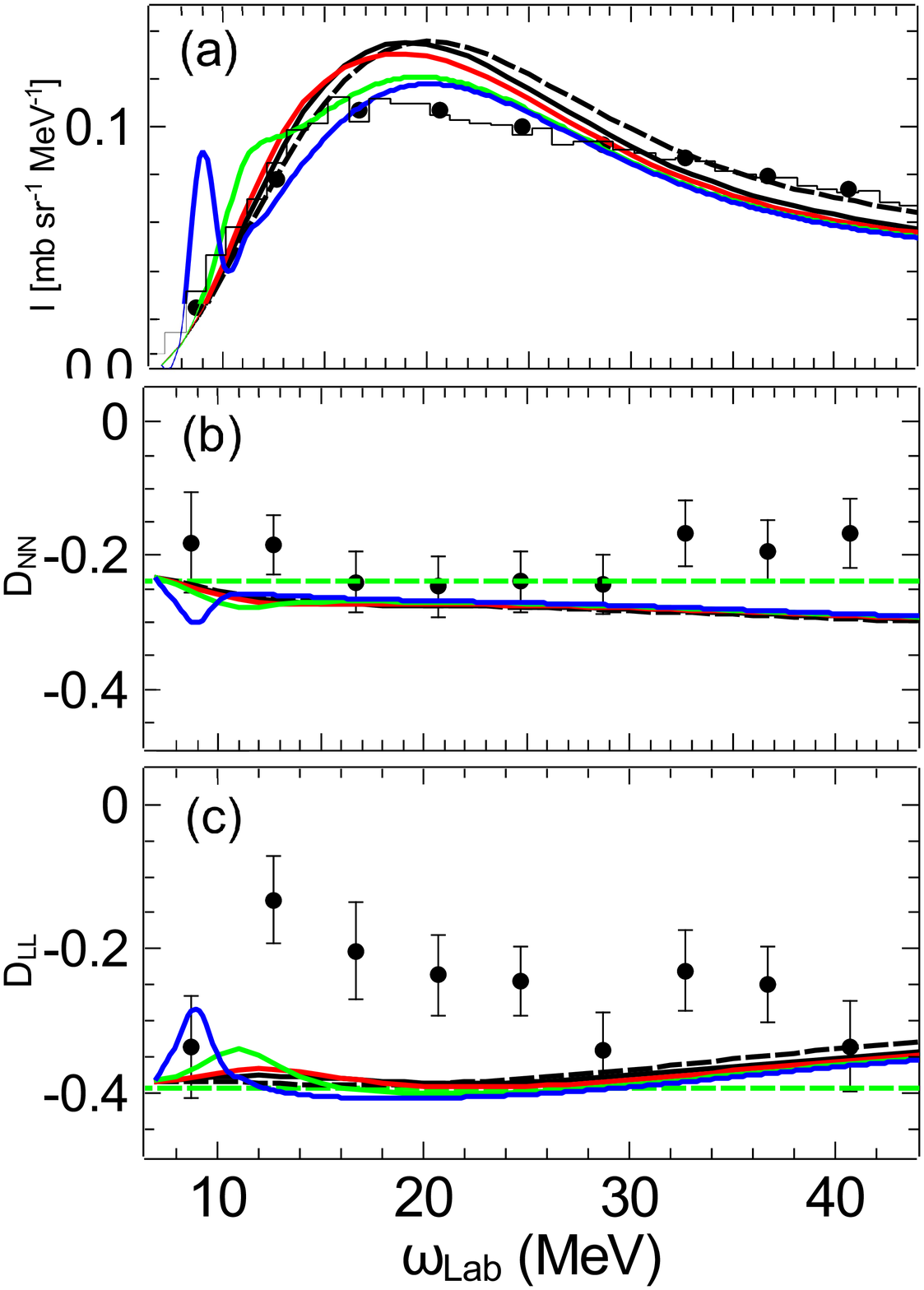}
\caption{
Same as Fig. \ref{fig:3He-pn-sig-DLL-DNN-2np}.
Calculations with AV18 are shown by dashed (black) lines.
Those with AV18 plus $V^{\mathrm{3NP}}(\frac{3}{2})$ that is active for  $J_0^{\pi}= \frac{1}{2}^-$ state taking $W_1(\frac{3}{2})=-36, -50, -70$, and $-90$ MeV, are shown by solid black, red, green, and blue lines, respectively.
\label{fig:3He-pn-sig-DLL-DNN-jp2}
}
\end{center}
\end{figure}

Fig. \ref{fig:3He-pn-sig-DLL-DNN-jp2} shows the observables calculated with $V^{\mathrm{3NP}}(\frac{3}{2})$ that is effective  only for $J_0^{\pi}= \frac{1}{2}^-$ state taking $W_1(\frac{3}{2})$ from -36 MeV to -90 MeV.
It looks that the 3NP with $W_1(\frac{3}{2})=-90$ MeV produces a resonance at $\omega_{Lab}=9$ MeV with a narrow width (about 2 MeV), which produces some visible effects on $D_{LL}(0^\circ)$ and $D_{NN}(0^\circ)$.
The width of the resonance is smaller than the reported in \cite{Wa08}.

\section{Summary}
\label{seq:Summary}

In this paper, I have presented calculations of the cross section and the polarization transfer coefficients, $D_{NN}$ and $D_{LL}$, in ${}^3\mathrm{He}(p,n)ppp$ $(\theta_{n}=0^{\circ})$ reaction with the spin-isospin response functions obtained for some realistic NN potential models.

The calculations have little NN potential dependence, and show a reasonable agreement with available experimental data, except that the energy-transfer dependence of $D_{LL}(0^\circ)$ is much smoother than the data.

Introductions of the attractive 3NP for the 3N$(T=\frac{3}{2})$ state suggested from the analysis of the $4n$ state as well as further strength enhanced 3NPs for $J_0^\pi=\frac{1}{2}^{-}$ state so as to produce a $3p$ resonance state  are examined.
But they cannot resolve the discrepancy. 

These results suggest that the curious energy-transfer dependence of the experimental $D_{LL}(0^\circ)$, which was the basis of the existence of the $3p$ resonance \cite{Wa08},  is not consistent with conventional models of the nuclear interaction, which indicates the need for further experimental studies of the reaction.

Also, a need for good knowledge of observables in $n(\vec{p},\vec{n})p$ at the very forward angles is stressed to reduce ambiguity in the calculation.

Finally, it is remarked that precise calculations of observables related to $3n$- or $3p$-system with theoretical models of the nuclear interactions are now available, which enables us to compare and then to study nuclear interactions whether a 3N resonance does exist or not.

\appendix
\section{Kinematics}

In this appendix, kinematical values related to ${}^{3}\mathrm{He}(p,n)ppp$ $(\theta_{n}=0^{\circ})$ reaction are summarized.

Let $T_{p}$ ($T_n$)  be the incident proton (outgoing neutron)  energy in Lab. system.
Masses of proton, neutron, and ${}^3$He are denoted by  $m_p$, $m_n$, and  $m_{{}^3\mathrm{He}}$, respectively.

\begin{itemize}

\item Total energy in Lab. system:
\begin{equation}
E_{tot,Lab} =  m_p + T_p + m_{{}^3\mathrm{He}}.
\end{equation}

\item The energy transfer and momentum transfer in Lab. system: 
\begin{subequations}
\begin{equation}
\omega_{Lab} = \left( m_p + T_p \right) - \left(m_n + T_n \right)
\label{eq:omega_lab}
\end{equation}
\begin{equation}
Q_{Lab} = K_{p} - K_{n},
\end{equation}
\end{subequations}
where
\begin{subequations}
\begin{equation}
K_{p} =\sqrt{ \left( m_p + T_p \right)^2 - m_p^2 }
\end{equation}
\begin{equation}
K_{n} =\sqrt{ \left( m_n + T_n \right)^2 - m_n^2 }.
\end{equation}
\end{subequations}

\item  Total energy of all four particles in c.m. frame of the initial $p$-${}^{3}\mathrm{He}$ system:
\begin{equation}
E_{tot,c.m.}  = \sqrt{E_{tot,Lab}^2 - K_{p}^2}
\end{equation}

\item Energy and momentum of the proton and energy of ${}^{3}\mathrm{He}$:  
\begin{subequations}
\begin{equation}
E_{p,c.m.} = \frac{E_{tot,c.m.}^2 + m_p^2 - m_{{}^3\mathrm{He}}^2}{2 E_{tot,c.m.} }
\end{equation}
\begin{equation}
k_i =  \sqrt{E_{p,c.m.}^2-m_p^2}. 
\end{equation}
\begin{equation}
E_{{}^3\mathrm{He},c.m.} = E_{tot,c.m.} - E_{p,c.m.} =  \frac{E_{tot,c.m.}^2 - m_p^2 + m_{{}^3\mathrm{He}}^2}{2 E_{tot,c.m.} }
\end{equation}
\end{subequations}

\item  Energies and momentum of the neutron, and energy of $3p$-system in the c.m. frame of the final $n$-$3p$ system:
\begin{subequations}
\begin{equation}
E_{n,c.m.}= \frac{E_{tot,c.m.}^2 + m_n^2 - (E_{3p,Lab}^2-Q_{Lab}^2)}{2 E_{tot,c.m.} }
\end{equation}
\begin{equation}
k_f = \sqrt{E_{n,c.m.}^2 - m_n^2},
\end{equation}
\begin{equation}
E_{3p,c.m.} = E_{tot,c.m.} - E_{n,c.m.}= \frac{E_{tot,c.m.}^2 - m_n^2 + (E_{3p,Lab}^2-Q_{Lab}^2)}{2 E_{tot,c.m.} },
\end{equation}
\end{subequations}
where
\begin{equation}
E_{3p,Lab} =  m_{{}^3\mathrm{He}} + \omega_{Lab}.
\end{equation}

\item Energy transfer and momentum transfer in the c.m. system: 
\begin{subequations}
\begin{equation}
\omega_{c.m.} 
= E_{p,c.m.} - E_{n,c.m.}
\end{equation}
\begin{equation}
Q_{c.m.} = k_i - k_f
\label{eq:Q_cm}
\end{equation}
\end{subequations}
\item The energy in the c.m. system of the final $3p$: 
\begin{equation}
E = \sqrt{E_{3p,c.m.}^2 - k_f^2} - 3 m_p.
\label{Eq:E_3p}
\end{equation}

\item The Kinematical factor in the expression,  Eq. (\ref{eq:I_dcs}), is given by
\begin{equation}
N_{K} = \left( \frac{2\pi}{\hbar}\right)^2 {\mu_i \mu_f} \frac{k_f}{k_i} 
\times \left(  \frac{K_{n}}{k_f} \right) \frac{dE}{d\omega_{Lab}}, 
\label{eq:N_K}
\end{equation}
where
\begin{subequations}
\begin{equation}
\mu_i = \frac{E_{p,c.m.} E_{{}^3\mathrm{He},c.m.}}{E_{p,c.m.} + E_{{}^3\mathrm{He},c.m.}},
\end{equation}
\begin{equation}
\mu_f = \frac{E_{n,c.m.} E_{3p,c.m.}}{E_{n,c.m.} + E_{3p,c.m.}},
\end{equation}
\end{subequations}
and
\begin{equation}
\frac{dE}{d\omega_{Lab}} 
= \frac{E_{tot,Lab}-E_{n,Lab}\frac{K_{p}}{K_{n}}}{E + 3 m_p}.
\label{eq:dE_domega_Lab}
\end{equation}

\end{itemize}


\end{document}